\documentclass{aastex63}
\usepackage{amsmath}
\usepackage{graphicx}
\usepackage{multirow}

\usepackage{xspace}

\begin{document}

\title{\rm  Proceedings of IAU Symposium 376 \\
Richard de Grijs, Patricia Whitelock and Marcio Catelan, eds.  \\ ~\\
{ \bf The Cepheid Extragalactic Distance Scale: Past, Present and Future }}

\author{Wendy L. Freedman }
\affiliation{Dept. of Astronomy \& Astrophysics, University of Chicago, 5640 S. Ellis Ave., Chicago, IL 60637, USA and \\ Kavli Institute for Cosmological Physics, University of Chicago, 5640 S. Ellis Ave., Chicago, IL 60637, USA}
\author{Barry F. Madore}
\affiliation{Carnegie Observatories, 813 Santa Barbara St., Pasadena CA 91101, USA and \\Dept. of Astronomy \& Astrophysics, University of Chicago, 5640 S. Ellis Ave., Chicago, IL 60637, USA}

\begin{abstract}
Cepheids have been the cornerstone of the extragalactic distance scale
for a century. With high-quality data, these luminous supergiants
exhibit a small dispersion in their Leavitt (period--luminosity)
relation, particularly at longer wavelengths, and few methods rival
the precision possible with Cepheid distances. In these proceedings,
we present an overview of major observational programs pertaining to
the Cepheid extragalactic distance scale, its progress and remaining
challenges. In addition, we present preliminary new results on
Cepheids from the {\sl James Webb Space Telescope} ({\sl JWST}). The
launch of {\sl JWST} has opened a new chapter in the measurement of
extragalactic distances and the Hubble constant. {\sl JWST} offers a
resolution three times that of the {\sl Hubble Space Telescope} ({\sl
  HST}) with nearly 10 times the sensitivity. It has been suggested
that the discrepancy in the value of the Hubble constant based on
Cepheids compared to that inferred from measurements of the cosmic
microwave background requires new and additional physics beyond the
standard cosmological model. {\sl JWST} observations will be critical
in reducing remaining systematics in the Cepheid measurements and for
confirming if new physics is indeed required. Early {\sl JWST} data
for the galaxy, NGC 7250 show a decrease in scatter in the Cepheid
Leavitt law by a factor of two relative to existing {\sl HST} data and
demonstrate that crowding/blending effects are a significant issue in
a galaxy as close as 20 Mpc.
\end{abstract}


\section{Introduction}

No review of the Cepheid distance scale would be complete without
recognition of the contribution by Henrietta Swan Leavitt. Leavitt's
1908 seminal paper 
(Leavitt 1908) on variable stars in the Large
(LMC) and Small (SMC) Magellanic Clouds established the relationship
between the period and luminosity of Cepheids. Subsequently, Leavitt
and Pickering (2012) 
published a
period--luminosity relation for 25 Cepheids in the SMC, finding a
nearly linear relationship between the logarithm of the luminosity and
the logarithm of the period. Leavitt and Pickering recognized that the
Magellanic Cloud Cepheids were likely at the same distance from Earth,
``... their periods are apparently associated with their actual
emission of light, as determined by their mass, density and surface
brightness.'' These critical insights laid the foundation for the
Cepheid distance scale, facilitating the measurement of distances to
astronomical objects. Leavitt passed away in 1921, and sadly did not
live to see the impact of her pioneering work. In 2008, a meeting was
held at Harvard University to commemorate the 100-year anniversary of
Leavitt's Cepheid discovery paper. An outcome of the meeting was the
recommendation to name the Cepheid period--luminosity relation the
`Leavitt Law' (Freedman \& Madore 2010).

Beginning in 1923, Edwin Hubble discovered Cepheids (Hubble 1925a,b, 1926)
galaxies M33, M31, and NGC 6822, and estimated the distances to these
galaxies employing Leavitt's period--luminosity relation. These and
subsequent measurements culminated in his discovery of the
relationship between galaxy distances and their recessional velocities (Hubble 1929),
and set the foundation for modern cosmology with
its implications for the expansion of the Universe.

Building upon a body of work characterizing the nature of variable
stars in Milky Way clusters (Bailey 1902, Joy 1949, Hogg 1955)
Baade (Baade 1956, Baade \& Swope 1963, 1965)
unambiguously established the existence of two populations of Cepheid
variables, resulting in a recalibration of Hubble's magnitude scale by
1.5 mag, or a factor of two in distance.

The next major advance in Cepheid studies came with
Wisniewski \& Johnson's (1968) photoelectric $UBVRIJKL$ observations
of Milky Way Cepheids. Their high-precision photometry exhibited two
major characteristics of Cepheid light curves: (1) the decrease in
amplitude moving from the ultraviolet to the near-infrared, and (2)
the shift in the relative phase of maximum light, also with increasing
wavelength.

The 1970s through 1990s saw the `factor-of-two' debate over the value
of H$_0$. In a series of ten papers entitled ``Steps Toward the Hubble
Constant'', published between 1974 and 1995, Allan Sandage and Gustav
Tammann (2006) 
presented evidence generally
favoring a value of H$_0 \sim 50$ km s$^{-1}$ Mpc$^{-1}$. In contrast, de Vaucouleurs (1993)
carried out his own calibration of the
distance scale, finding values closer to 100 km s$^{-1}$
Mpc$^{-1}$. An early detailed review of these programs was given by Hodge (1981)
Hodge optimistically summarized the controversy:
``Eventually, I assume, when sufficiently reliable measurements are
possible ... there will be some number, such as H$_0$, that everyone
will agree on, within some reasonable and understood limits.'' It
would be another 20 years before the {\sl Hubble Space Telescope}
({\sl HST}) Key Project team (Freedman et al. 2001) 
measured a value
of H$_0$ to 10\% accuracy; a number that has stood the test of time.

The accurate measurement of Cepheid distances remains non-trivial.
One of the main challenges is distinguishing Cepheids from fainter
stars, both resolved and unresolved, that contribute to the overall
background of the host galaxy. Avoiding/minimizing the effects of
crowding and confusion is crucial for the successful identification
and measurement of Cepheids in galaxies. A second challenge is the
correction for the presence of dust, which both scatters and absorbs
radiation. Cepheids are relatively young stars that have not had time
to diffuse far from the locations where they were formed; hence they
are found in the dusty disks of spiral (or irregular) galaxies. Third,
the luminosity of a Cepheid may have a dependence on the abundance or
metallicity of the star.

\section{Advantages of the Near-Infrared}

\begin{figure}
\centering
\includegraphics[width=\columnwidth]{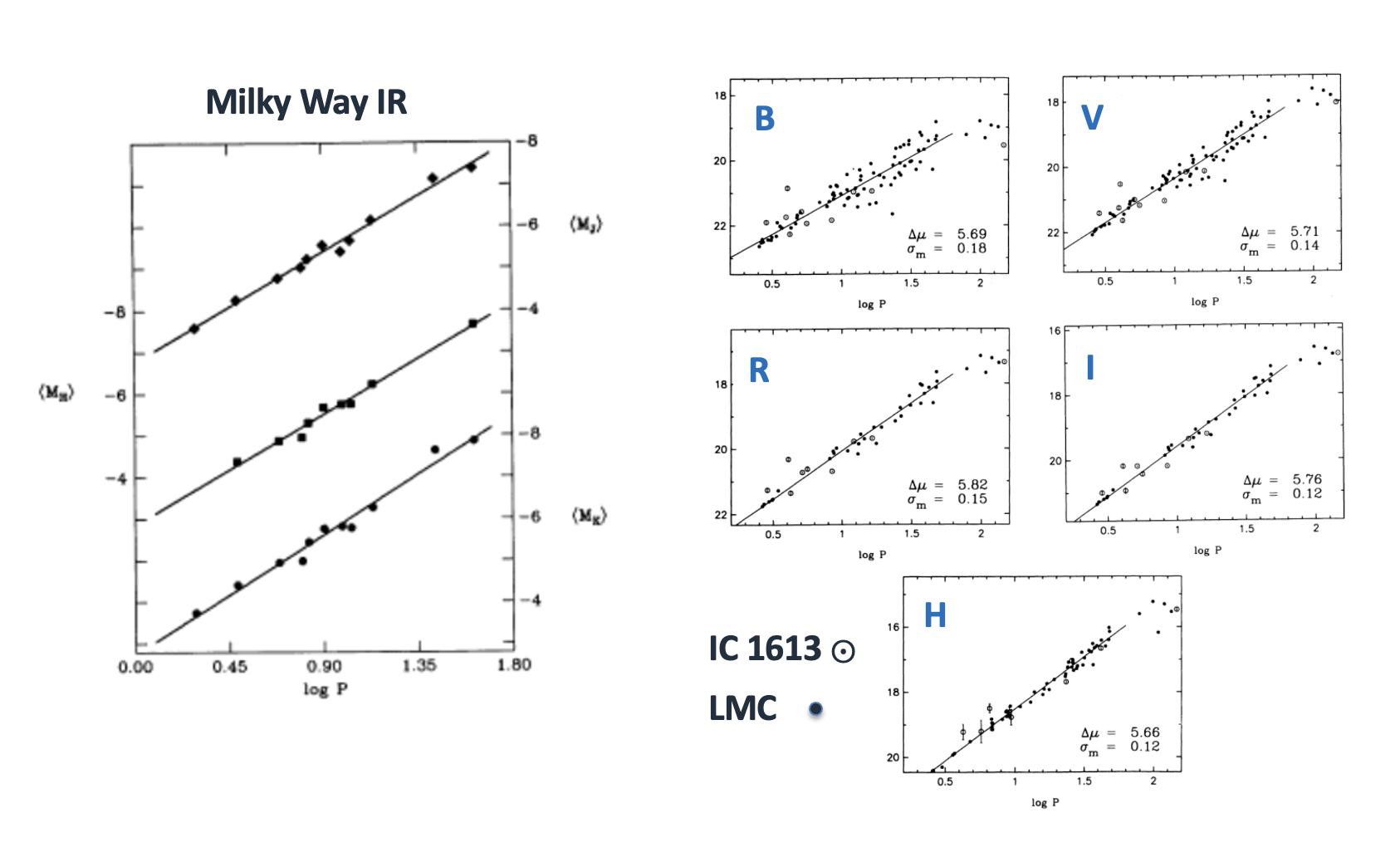}
\caption{ Left panel: Leavitt laws at JHK wavelengths (McGonegal et al. 1983)
demonstrating the power of infrared photometry for the extragalactic distance scale. Right panel:  Leavitt laws at BVRI wavelengths based on CCD photometry (Freedman 1988)
(open circles); filled circles illustrate photoelectric photometry for LMC Cepheids.    }
\label{fig:McG+WLF}
\end{figure}

The recognition of the applicability of near-infrared observations of
Galactic Cepheids to the distance scale did not occur immediately,
despite the early near-infrared observations made by Wisniewski \& Johnson (1968)
The first application was by McGonegal et al. (1982)
(see Figure \ref{fig:McG+WLF},
left), who elucidated the advantages of the near-infrared, most
specifically the decreased sensitivity to reddening and the decrease
in the observed width of the Leavitt law, even with only single-epoch
observations. The decrease in the width of the period--luminosity
relation is primarily due to the decrease in sensitivity of the
infrared surface brightness to the intrinsic temperature width of the
instability strip. Near-infrared observations allowed for accurate
distance determinations even with single, random-phase observations of
Cepheids whose periods had already been measured. These single-phase
measurements were comparable in accuracy to complete time-averaged
magnitudes derived from many nights of observations obtained at blue
wavelengths.

\section{The Era of CCDs and Infrared Arrays}

With a wide wavelength sensitivity range, spanning from the $B$ band
(0.45 $\mu$m) to the $I$ band (0.8 $\mu$m), CCD detectors allowed, for
the first time, the possibility of high-resolution, panoramic, digital
data for the extragalactic distance scale 
(Freedman 1984).
With
CCDs one could perform local sky subtraction and point-spread-function
fitting to obtain accurate magnitudes and colors. In addition, CCDs
enabled the handling of crowding and confusion errors at the
one-square-arcsecond level. Furthermore, the large wavelength coverage
of CCDs allowed the explicit and direct determination of reddening, by
applying the interstellar extinction law as a function of wavelength,
and eliminating the need to rely solely on foreground estimates and/or
assuming negligible additional reddening (in the host galaxy), as many
earlier studies had done. The approach, introduced by Freedman (1988)
for determining total line-of-sight extinctions
and true distance moduli of extragalactic Cepheids' multiwavelength
data (e.g., see Figure \ref{fig:McG+WLF}, right) applied to
single-epoch observations of Cepheids in IC 1613, M31, M33, and NGC
300, among others. In brief, the method involves determining
differential apparent moduli by scaling them with respect to the
corresponding LMC period--luminosity relations and assuming that the
(observed) variation with inverse wavelength can be attributed to
selective absorption.

\medskip\medskip\medskip

\subsection{Metallicity Effects}

In addition to reddening, nailing down the metallicity dependence of
Cepheids is also critical for application of the Leavitt law.
Metallicity can affect the internal structure of a Cepheid: a higher
metallicity results in a higher opacity, which can affect the
temperature, radius, and luminosity of the star. Line blanketing,
controlled by the abundance of metals in the atmospheres of Cepheids,
can also result in a redistribution of light as a function of
wavelength. The atmospheric effects are predicted to be greater at
optical wavelengths than in the infrared.

The availability of CCDs also allowed the first empirical test of
metallicity on the Cepheid period--luminosity
relation. Freedman \& Madore (1990)
obtained $BVRI$ CCD observations
of Cepheids in M31 and estimated the mean reddenings and true distance
moduli at three different radial positions (3, 10, and 20 kpc) in the
disk of M31, for which a metallicity gradient of a factor of five had
been measured from H{\sc ii} regions. They found evidence for a mild
gradient of $-0.2 \pm 0.2$ mag dex$^{-1}$. However, given the
uncertainties, the value was also consistent with no measurable
effect.

Twenty-five years later, there still remains uncertainty in the
metallicity coefficient Ripepi et al. (2020)
A number of recent
studies have found a range in the metallicity coefficient from $-0.5$
to 0 mag dex$^{-1}$ (Udalski et al. 2001, Riess et al. 2016, Riess et al. 2021, Breuval et al. 2021, Ripepi et al. 2021, Ripepi et al. 2022).
  Generally, the effects of
metallicity are found to be larger in the optical than in the
near-infrared, although in contrast, the recent study by Breuval et al. (2022)
finds nearly as large an effect in the
near-infrared as in the optical.

\section{The Hubble Space Telescope Key Project}

The goal of the Hubble Key Project 
(Freedman et al. 2001) 
was to extend
the Cepheid extragalactic distance scale to greater distances than
possible from the ground, tie in to several independent methods for
measuring distances beyond the Cepheids, and to ultimately measure the
Hubble constant to an overall accuracy of 10\%. To achieve these
goals, the Key Project team used the {\sl HST} to observe Cepheids in
18 galaxies spanning a range of distances. The observations included
both wide-field and deep imaging in the F555W and F814W filters, with
F555W ($V$ band) allowing for the discovery, characterization, and
measurement of the periods, amplitudes, and mean magnitudes of the
Cepheids; and using the longer-wavelength F814W ($I$ band) to form a
color, thereby providing the means to correct for the effects of
extinction and absorption by dust.

\begin{figure}
\centering
\includegraphics[width=8cm]{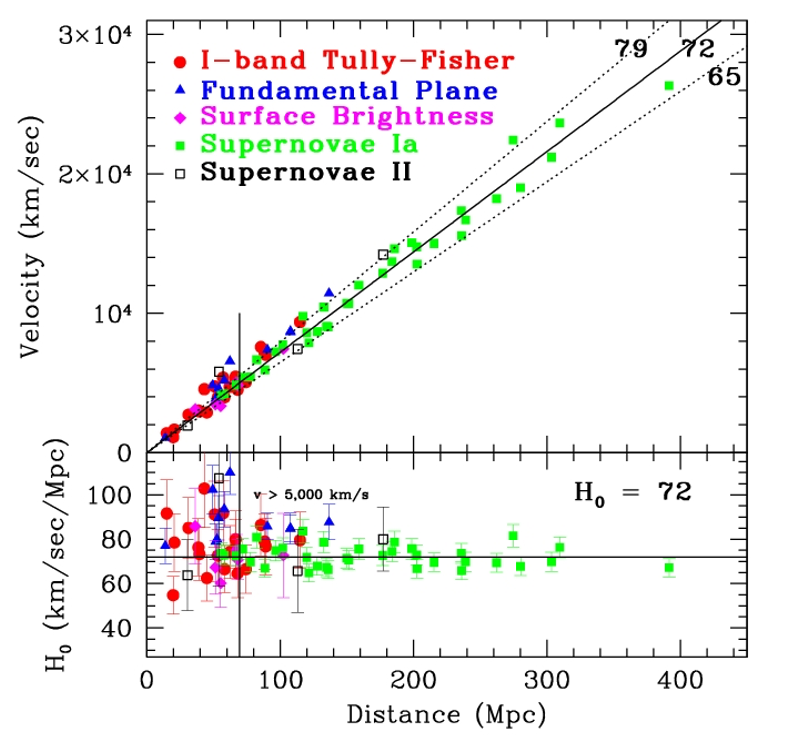}
\caption{Hubble diagram showing the final results from the {\sl HST}
  Key Project (Freedman et al. 2001)
  In the top panel velocity is
  plotted versus distance for the five secondary distance methods
  labeled in the plot. The solid line is for a slope of 72, flanked by
  dotted lines indicating a 10\% uncertainty. The bottom panel shows
  the residuals for a value of H$_0 = 72 \pm 3$ (stat) $\pm 7$ (sys) km s$^{-1}$
  Mpc$^{-1}$.}
\label{fig:HoKP}
\end{figure}

These measurements then enabled the calibration of secondary distance
indicators (including Type Ia and II supernovae, the Tully--Fisher
relation, surface brightness fluctuations, and the $D_n$--$\sigma$
relation; see Figure \ref{fig:HoKP}. The final value of the Hubble
constant was measured as H$_0 = 72 \pm 3$ (statistical) $\pm 7$ (systematic) km s$^{-1}$ Mpc$^{-1}$, with
an estimated uncertainty of about 10\%. This result provided early
evidence for a Universe that is accelerating, consistent with the
presence of dark energy, and consistent with the ages of globular
clusters. And it resolved the longstanding factor-of-two discrepancy
in the distance scale.

\section{NICMOS and Ground-Based Near-Infrared Observations of Cepheids}

\subsection{Early NICMOS Observations}

\begin{figure}
\centering
\includegraphics[width=8cm]{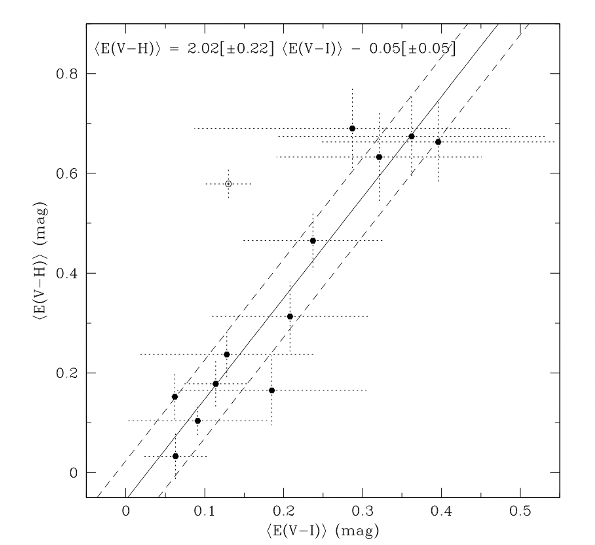}
\caption{A comparison of reddening values measured via $E(V-H)$ with
  those from $E(V-I)$ from Macri et al. (2001). The good agreement provides evidence that a
  standard extinction law is a reasonable approximation to correct for
  the dust in these galaxies.}
\label{fig:macri}
\end{figure}

Following on the {\sl HST} project, near-infrared observations of 14
nearby galaxies were obtained with {\sl HST}/NICMOS (Macri et al. 2001).
Distance moduli were obtained by combining
single-epoch $H$-band data with the $VI$ bands. These data provided
support for the assumption of the universality of the standard (Milky
Way) reddening law adopted by the Key Project (see Figure
\ref{fig:macri}), by demonstrating the excellent correspondence
between $E(V-H)$ and $E(V-I)$.

An additional goal of the NICMOS program was to undertake another test
for the metallicity effect by observing two fields in the relatively
face-on galaxy M101, a test that had earlier been undertaken by the
Key Project team. The inner and outer M101 fields were found to have a
large difference in distance modulus (amounting to $+0.41 \pm 0.11$
mag at $H$ and $+0.34 \pm 0.09$ mag at $J$, respectively). If
interpreted in terms of a metallicity effect, this implies a
dependence of order 0.6 mag dex$^{-1}$, much higher than theory
predicts for the near-infrared. Simulations, as well as artificial
star tests showed that crowding/blending effects were likely an issue
for these data, explaining some (but not all) of this large
difference.

These results provided a cautionary note regarding the difficulty of
measuring Cepheids in the near-infrared in all but the nearest
galaxies, and the need to avoid their inner (high-surface-brightness)
regions. While the effects of dust are reduced at increasing
wavelengths, the background contamination by red stars (red giants and
even brighter asymptotic giants) increases substantially. Thus, the
advantages of moving to the near-infrared to mitigate extinction
effects were countered by the systematic disadvantages imposed by
crowding and blending.

\medskip
\medskip

\subsection{Ground-Based Near-Infrared Observations of the LMC }

\begin{figure}
\centering
\includegraphics[width=10cm]{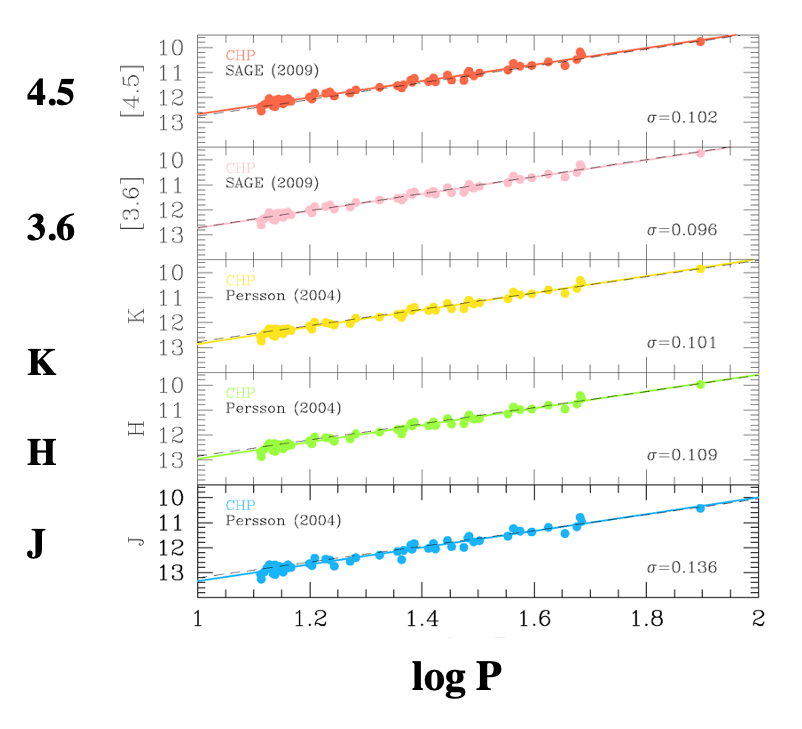}
\caption{Leavitt laws for near-infrared ($JHK$) and mid-infrared (3.6
  and 4.5 $\mu$m) for Cepheids in the LMC. }
\label{fig:LMC_IR_PLs}
\end{figure}

From the ground, an extensive and uniform near-infrared ($JHK$) study
of 92 Cepheids in the LMC was undertaken by Persson et al. (2004), 
with $JHK$ sampling obtained at about 22 phase
points for each star. The dispersion about the Leavitt law was found
to be on the order of $\pm$0.1 mag for all three wavebands (see Figure
\ref{fig:LMC_IR_PLs}). Fitting for a tilt to the LMC, the dispersions
in the near-infrared period--luminosity--color relations drop to
$\sim$0.08 mag, showing again the immense power of the infrared in
cases where crowding/blending effects were intentionally minimized and
where the signal-to-noise is high. The high quality of these data is
apparent when cross-comparing the positions of the individual Cepheids
in these period--luminosity relations. The deviations from the mean
period--luminosity relations are highly correlated across the various
bandpasses, demonstrating intrinsic properties of the Cepheids and/or
location within the LMC, that are not blurred out by photometric
errors.

\section{The Spitzer Space Telescope: Mid-infrared Observations of Cepheids}

Using the IRAC camera on the {\sl Spitzer Space Telescope} during its
extended, warm mission, well-sampled mid-infrared light curves were
obtained for 80 long-period LMC Cepheids 
(Scowcroft et al. 2011)
falling
in the period range $0.8 < \log(P) < 1.8$ [days]. These observations
provided a well-measured slope of the period--luminosity relation
based on the time-averaged 3.6 $\mu$m data. Time-averaged 3.6 $\mu$m
data were also obtained for 10 high-metallicity Milky Way Cepheids
with independently measured {\sl HST} trigonometric parallaxes (Monson et al. 2012).
The 3.6 and 4.5 $\mu$m Leavitt laws for the LMC
can be seen in Figure \ref{fig:LMC_IR_PLs}. The 3.6 $\mu$m data
resulted in a high-precision, reddening-corrected distance to the LMC
of $18.477 \pm 0.033$ (systematic) mag. Freedman et al. (2012)
applied the {\sl Spitzer} calibration to the Key
Project galaxy sample, yielding a value of H$_0 = 74.3$ km s$^{-1}$
Mpc$^{-1}$, with a systematic uncertainty of $\pm$ 2.1 (systematic) km
s$^{-1}$ Mpc$^{-1}$. The {\sl Spitzer} program reduced the systematic
uncertainty in H$_0$ over that obtained by the {\sl Hubble Space
  Telescope} Key Project by more than a factor of three.

\section{The Araucaria Project}

For 20 years, the Araucaria Project has been carrying out a program to
observe a number of types of variable stars in nearby galaxies at both
optical and near-infrared wavelengths (Pietrzynski et al. 2019)
They
have monitored Cepheid variables, RR Lyrae stars, and eclipsing
binaries, as well as blue supergiants, tip of the red-giant branch
(TRGB) stars, and carbon stars. The Araucaria Project team has
measured Cepheid distances to galaxies in the Local Group (LMC, SMC,
IC 1613, NGC 6822, M33), the Sculptor Group (NGC 247, NGC 300, NGC
7793), providing fundamental distances and tests of Cepheid
metallicities (Udalski et al. 2001).

\medskip
\medskip

\section{The SH0ES Program}

The SH0ES (Supernovae, H$_0$, for the Equation of State) project,
initially aimed at the study of dark energy, now has the objective of
determining H$_0$ through the calibration of the extragalactic
distance scale using Cepheid variables (Riess et al. 2009, 2012, 2016, 2022) 
  Riess and colleagues report {\sl HST}
F555W, F814W, and F160W observations of Cepheid variables in host
galaxies of Type Ia supernovae (SNe~Ia) used to calibrate
H$_0$. Reddening corrections are obtained using a small number of (2
to 4) random-phase observations in the F814W and F555W bands, with the
distances coming mainly from $\sim$11 observations taken in the F160W
($H$) band. The scatter in the F160W period--luminosity relations is
generally of order 0.4--0.5 mag, i.e., a factor of four or so greater
than the intrinsic dispersion observed in the uncrowded sample of
Cepheids in the LMC. The zero-point calibration is set by geometric
{\sl Gaia} early Data Release 3 (EDR3) parallaxes, masers in the
galaxy NGC 4258, and detached eclipsing binaries in the LMC. The most
recent SH0ES paper quotes a 1\% uncertainty with H$_0 = 73.04 \pm
1.04$ km s$^{-1}$ Mpc$^{-1}$, based on a sample of 42 galaxies with
distances in the range from 7 out to 80 Mpc.

\section{An Independent External Check of Cepheid Distances: Tip of the Red Giant Branch}

Few methods for measuring nearby distances rival the Cepheids in
precision, accuracy, or numbers of galaxies for which the measurement
can be made. A notable exception is the TRGB method (Lee et al. 1993, Mould \& Sakai 2009, Hoyt 2023).
The TRGB method also has a number of
advantages relative to the Cepheid Leavitt law (Freedman et al 2019, Freedman 2021).
It can be applied in the halos of galaxies where the
effects of dust are generally negligible and where the surface
brightness is about 100 times lower than in the disks, so that
crowding/blending effects are not severe. It is a much easier
practical issue to measure the luminosity of a halo TRGB star than it
is to disentangle Cepheids from nearby neighbors in the disk. In
addition, metallicity effects are much smaller for the TRGB (and can
be directly calibrated), unlike in the case of Cepheids. Thus, beyond
providing an independent calibration of H$_0$ (Freedman et al. 2019),
the TRGB provides an excellent external constraint on the combined
systematics of both methods.

\begin{figure}
\centering
\includegraphics[width=\columnwidth]{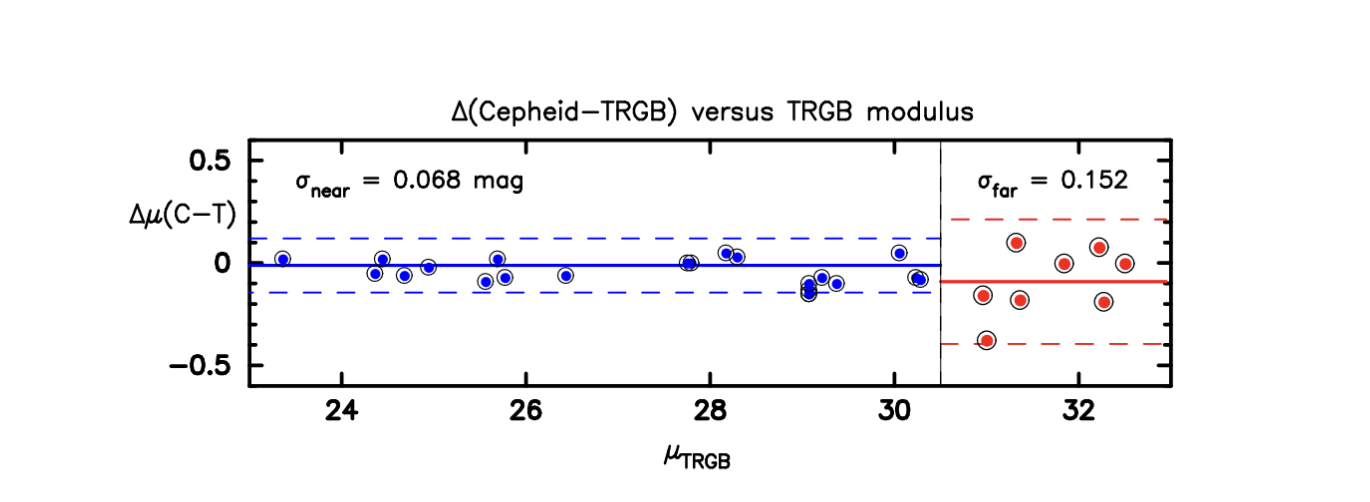}
\caption{Comparison of published Cepheid and TRGB distances in the
  sense of (Cepheid $-$ TRGB). The blue/red solid lines show the mean
  offset for galaxies with distance moduli closer/farther than 30.5
  mag. The dashed lines indicate the 1$\sigma$ scatter for the near
  and far samples, which are also labeled in the plot. The agreement
  between the Cepheid and TRGB distances is excellent, particularly
  for the nearby sample.}
\label{fig:ceph_trgb}
\end{figure}

In Figure \ref{fig:ceph_trgb} we show a comparison of distance moduli
for Cepheids and the TRGB. The agreement between the two distance
scales is excellent, with $\Delta\mu$(Cepheid $-$ TRGB) $= -0.026 \pm
0.014$ mag. In this case, the zero point for the TRGB calibration is
$M_I = -4.05$ mag. The scatter (expectedly) increases with increasing distance and photometric errors.

\begin{figure}
\centering
\includegraphics[width=12cm]{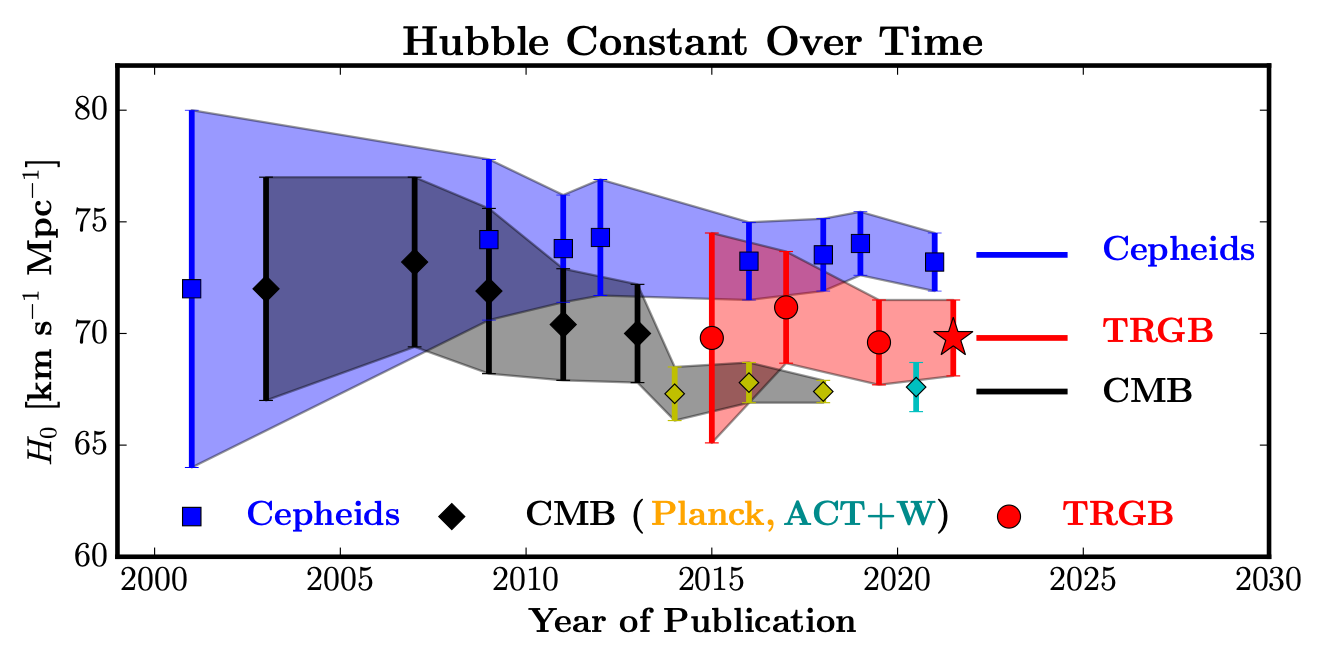}
\caption{Measurements of H$_0$ from the {\sl HST} Key Project through
  to today based on Cepheids (blue), the TRGB (red), and the CMB ({\sl
    Planck}: yellow; ACT + WMAP: turquoise).}
\label{fig:Hosummary}
\end{figure}

In Figure \ref{fig:Hosummary} we show an update to a recent comparison
of the values of H$_0$ obtained from Cepheids (blue) and the TRGB (red); both tied into measurements of SNe~Ia (Freedman 2021)
and that inferred from observations of the cosmic microwave background (CMB; Planck Collaboration et al. 2020, black).
The Cepheid and TRGB H$_0$ values
agree to within 2$\sigma$ of their quoted uncertainties; the TRGB
measurements are in better agreement with the CMB value to within
1$\sigma$.

\section{Remaining Issues for the Cepheid Distance Scale}

As previously noted, several challenges are still outstanding in
measuring Cepheid distances, resulting from the fact that Cepheids are
located in regions of star formation within the disks of galaxies,
hence in dusty and crowded regions, particularly in their inner disks.
The amount (and even the sign) and the wavelength dependence of the
metallicity contributions to Cepheid luminosities, within and between
galaxies, has yet to converge (Ripepi et al. 2020, Breuval et al. 2021,  da Silva et al. 2022, Owens et al. 2023).

\begin{figure}
\centering
\includegraphics[width=12cm]{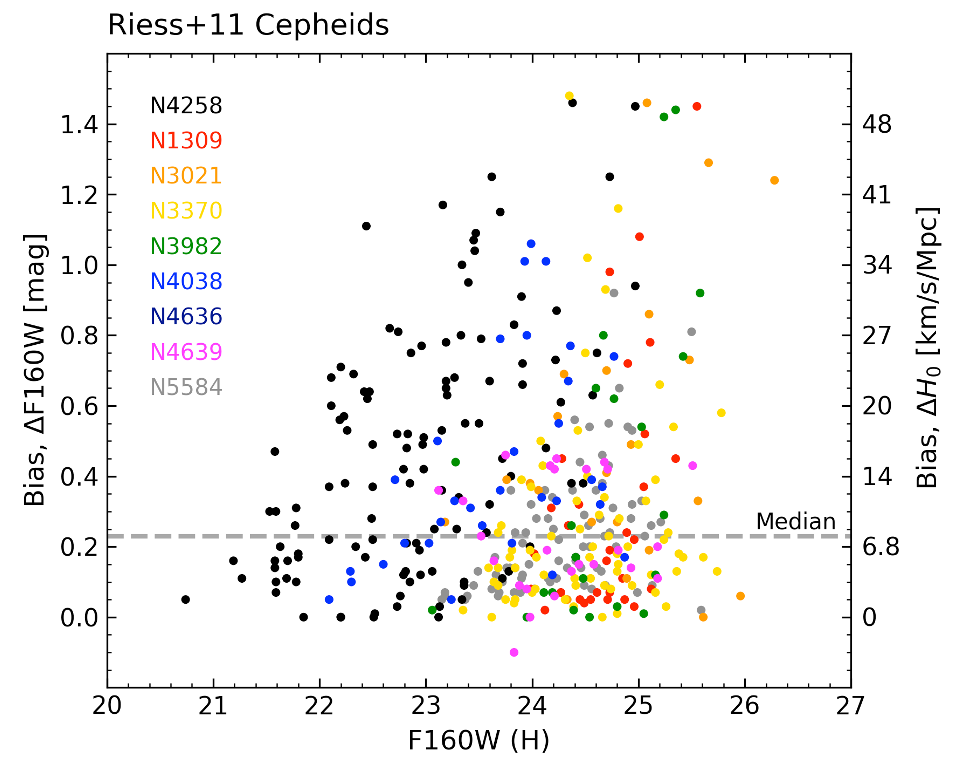}
\caption{A measure of the bias in Cepheid photometry from Table 2 as
  published by Riess et al. (2011).
  The bias is determined from
  artificial star tests for each Cepheid's environment and is a
  statistical correction. The bias can be quite significant and ranges
  from 0 to 1.5 mag (a factor of up to four in luminosity). The median
  bias is 0.24 mag, corresponding to $\Delta$H$_0 = 7$ km s$^{-1}$
  Mpc$^{-1}$. The galaxies in this earlier study are amongst the
  closest in the more recent sample of 42 galaxies in the Riess et al. (2022).}
\label{fig:crowding}
\end{figure}

Both the {\sl HST} Key Project and the SH0ES group approach the issue
of crowding/blending of Cepheids with the use of artificial star
tests. Figure \ref{fig:crowding} (courtesy of I. Jang) shows the F160W
crowding-induced bias found for Cepheids observed in nine of the
closest SNe~Ia host galaxies (Riess et al. 2011). 
As can be seen, the
bias can be quite large, reaching over 1 mag, with a median correction
of $\sim$0.25 mag, which corresponds to a 10\% difference in H$_0$ of
$> 7$ km s$^{-1}$ Mpc$^{-1}$ (as shown to the right of the plot). With
the (unexplained) scatter in the F160W period--luminosity relation
being as large as 0.4--0.5 mag, accurate crowding corrections are
critical. And within that scatter, there remains the potential for a
hidden systematic effect that may be difficult to identify and account
for.

It is important to note that on average, for similar types of
galaxies, crowding effects will become more severe with increasing
distance. Sixty percent of the Riess et al. (2022) sample of galaxies
in which Cepheids have been discovered lie at greater distances than
20 Mpc, and 25\% of the total sample lies beyond 40 Mpc. At a distance
of 40 Mpc, four times the area will be contained within a given
pixel. For the most distant SH0ES galaxy at 80 Mpc, 16 times the area
will be covered. As the need for percent-level accuracy has grown in
importance, and given the level of crowding for Cepheids in a galaxy
at a distance of only 20 Mpc (see the {\sl JWST} observations in the
following section), it remains important to demonstrate that crowding
effects do not produce a systematic bias in the photometry and, hence,
the distance measurements for these more distant galaxies observed
with {\sl HST}.

\section{The Future: Cepheids and JWST}

While measurements of the Cepheid distance scale have improved since
the time of the Key Project, great strides have been made in
measurements of fluctuations in the temperature and polarization of
the microwave background (Planck Collaboration 2018).
The current controversy
over the value of H$_0$ now requires pushing the determination of
local values to limits approaching 1\% accuracy, in order to be of
comparable accuracy to the CMB measurements. As we have seen,
systematic errors have played, and continue to play, a dominant role,
keeping more accurate measurement of H$_0$ a challenge.

Much like the 1970s and 1980s, when local distance measurement
differences remained at an impasse, it is clear once again that
improvements in technology will be required to settle the current
differences, at the few-percent level, for the Cepheid extragalactic
distance scale. We describe here a new program using the {\sl James
  Webb Space Telescope} ({\sl JWST}), aimed at addressing the current
systematics in the {\sl HST} measurements of Cepheid distances. The
program is actually three-pronged and includes also the measurement of
distances to nearby galaxies using not only Cepheids, but also TRGB
stars and carbon/JAGB stars. In this review, however, we describe only
the Cepheid portion of this program.

Because of the greater sensitivity of the surface brightness to
temperature at bluer (optical) wavelengths, the amplitudes of Cepheid
variables are larger than at longer wavelengths. Thus, its blue
sensitivity and high spatial resolution (compared to the ground) made
{\sl HST} an ideal facility for the {\it discovery} of Cepheid
variables (Freedman et al. 2001, Riess et al. 2016).

Two key features now make {\sl JWST} the optimal telescope for
addressing the {\it accuracy} of measurements of H$_0$: its red
sensitivity and higher spatial resolution (compared to {\sl HST}, and
certainly far exceeding the spatial resolution and sensitivity of {\sl
  Spitzer}). The extinction is significantly lower: for example, $A_J$
is smaller by a factor of four relative to the visual extinction,
$A_V$ (Cardelli et al. 1989).
For the F115W ($J$-band) filter on {\sl
  JWST}/Near-Infrared Camera (NIRCAM; Rieke et al. 2023),
  the
sampling resolution is four times that of {\sl HST}/Wide-Field Camera
3 (WFC3) F160W ($H$ band), with a FWHM of 0.04 arcsec versus 0.15
arcsec. With $\sim 4 \times$ better resolution than {\sl HST/WFC3},
crowding effects are reduced by more than an order of magnitude in
flux. This is especially important because, in the near-infrared,
contamination of the Cepheids by red giant and asymptotic giant branch
stars exacerbates crowding effects relative to optical wavelengths.

As part of the Chicago Carnegie Hubble Program (CCHP), we have been
awarded time in {\sl JWST} Cycle 1 (JWST-GO-1995: PI W. L. Freedman,
Co-PI B. F. Madore) for a program designed to address specifically
known systematic effects: extinction and reddening by dust,
metallicity effects, and crowding/blending of stellar images. We are
obtaining NIRCAM $J$-band data and, simultaneously, F356W
(mid-infrared) observations of 10 nearby SNe~Ia-host galaxies with
known Cepheids. For calibration purposes, we are also observing NGC
4258, a galaxy with a geometric distance derived from the motions of
nuclear H$_2$O megamasers. In addition to the {\sl JWST} results, our
group is undertaking a complete re-analysis (Owens et al. 2023) 
of the
archival optical and near-infrared {\sl HST} data obtained by the
SH0ES group (Riess et al. 2016).

\begin{figure}
\centering
\includegraphics[width=\columnwidth]
{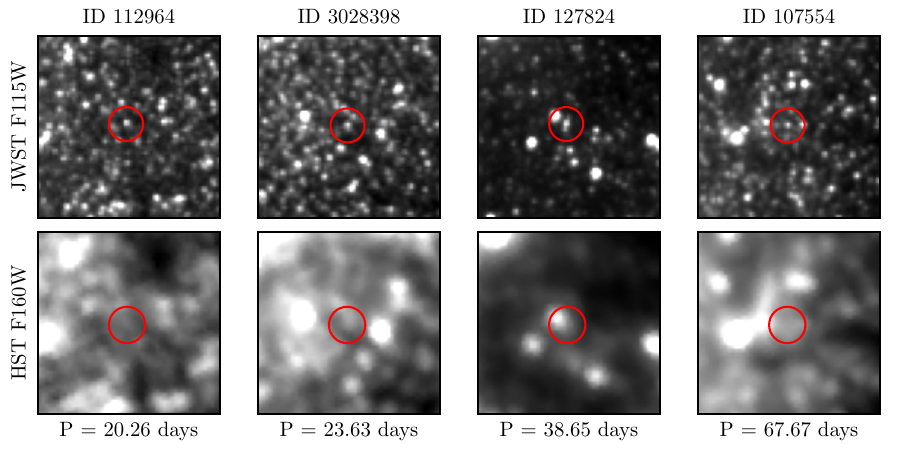}
\caption{Four Cepheids in NGC 7250 discovered as part of the SH0ES
  project (bottom row: {\sl HST} F160W/$H$-band exposures; top row:
  {\sl JWST} F115W/$J$ band). Each postage-stamp image is $2 \times 2$
  arcsec on a side. The red circles are centered at the position of
  the Cepheid determined from the {\sl HST} optical (F350LP white
  light) photometry. It is immediately evident that the crowding for
  these Cepheids is quite severe and the signal-to-noise ratio (SNR)
  for the $H$-band data tends to be low, ranging from 1 to 23. In
  contrast, the SNR for the $J$-band data ranges from 36 to 121. On
  average, the {\sl JWST} data have almost an order of magnitude
  greater SNR, and a four times better angular resolution, allowing
  the Cepheids to be distinguished clearly from the background.  }
\label{fig:CephPost}
\end{figure}

Here, we present some preliminary results from our program for the
nearby galaxy NGC 7250, host to SN2013DY, located at a distance of
about 20 Mpc. Figure \ref{fig:CephPost} illustrates cut-out images of
four Cepheids in NGC 7250 from (Owens et al. 2023).
Shown are images at F115W (from {\sl JWST}; top
panel) and F160W (from {\sl HST}; bottom panel), which illustrate the
superb resolution and the power of {\sl JWST}. As can be seen in the
{\sl HST} data, many of the Cepheid candidates are crowded by nearby
neighbors.

\begin{figure}
\centering
\includegraphics[width=\columnwidth]{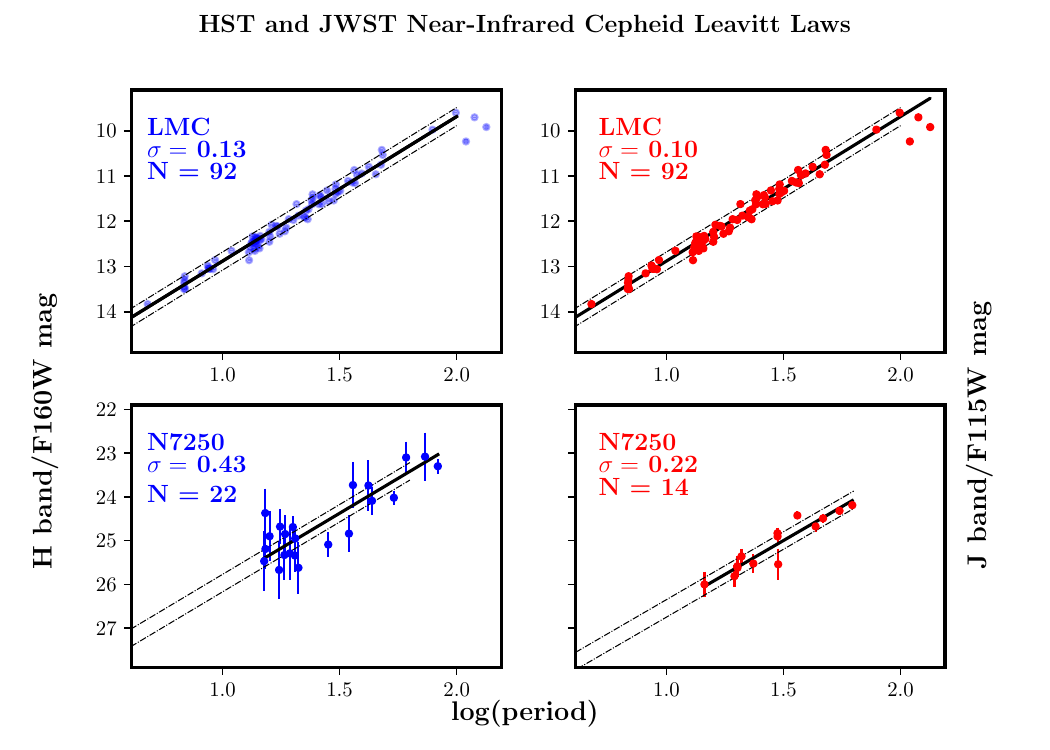}
\caption{Period--luminosity relations for Cepheids in the LMC (top
  row) and NGC 7250 (bottom row). The left-hand panels are for
  ground-based $H$ band for the LMC (top) and the similar F160W filter
  (from {\sl HST}) for NGC 7250 (bottom). The right-hand panels are
  similar, but for ground-based $J$ and the corresponding F115W filter
  (from {\sl JWST}). The {\sl JWST} data are plotted on an arbitrary
  magnitude scale. The scatter about the period--luminosity fit in
  each filter is labeled in each plot.}
\label{fig:Cepheid_PL}
\end{figure}
\vfill\eject
In Figure 9, we show the Leavitt laws for a sample of Cepheids in the
LMC (Persson et al. 2004)
and in NGC 7250 (Owens et al. 2023).
The scatter in the {\sl JWST} F115W data for NGC 7250 is a factor of two
smaller than in the SHoES F160W data; i.e., the improved resolution
and higher signal-to-noise ratio of the {\sl JWST} data results in a
lower variance ($\sigma^2$) for the F115W relation by almost a factor
of four. This is all the more remarkable since the $J$-band data are
single-phase observations only, while the {\sl HST} observations have
been corrected to mean light. The {\sl HST} data exhibit more than
three times the scatter of the $H$-band data for the LMC, the latter
of which reflects the expected scatter for that band, as exemplified
by the LMC data.

\section{Summary}

The accuracy of the Cepheid distance scale has continued to improve
over the century during which it has been used to measure the
distances to nearby galaxies and set the scale for the determination
of H$_0$. Still, challenges remain in overcoming systematic
uncertainties. Many of these challenges will be overcome with new
capabilities provided by the {\sl JWST}.

New {\sl JWST} data for the nearby galaxy NGC 7250 already demonstrate
that (1) many of the Cepheids observed with {\sl HST/WFC3} are
significantly crowded (and biased to brighter apparent magnitudes) by
nearby neighbors. A re-analysis of the SH0ES optical data, then
coupled with the new high-resolution and higher signal-to-noise {\sl
  JWST} F115W data, leads to significantly reduced effects of crowding
and smaller photometric uncertainties. (2) These improvements result
in a factor of two lower scatter in the near-infrared Leavitt law for
{\sl JWST} F115W compared with {\sl HST} F160W, even with single-epoch
F115W {\sl JWST} photometry.

The galaxies in our {\sl JWST} CCHP program sample have all been
selected to have with distances $\lesssim$20 Mpc, close enough to
minimize crowding effects. As for the case of NGC 7250
presented here, these data will be combined with a re-analysis of the
SH0ES {\sl HST} optical data for the Cepheids. TRGB, carbon star, and
Cepheid distances to the same sample of galaxies being observed as
part of the CCHP will allow measurement of three independent distances
to each of these galaxies. These new results already illustrate the
power of {\sl JWST} to improve the measurement of extragalactic
distances, and specifically, to address remaining systematics in the
determination of H$_0$.

Ultimately, these data will allow us to provide an answer to one of
the most important problems in cosmology today: Is there new
fundamental physics required beyond standard $\Lambda$ Cold Dark
Matter?
\vfill\eject
\begin{figure}
\centering
\includegraphics[width=\columnwidth]{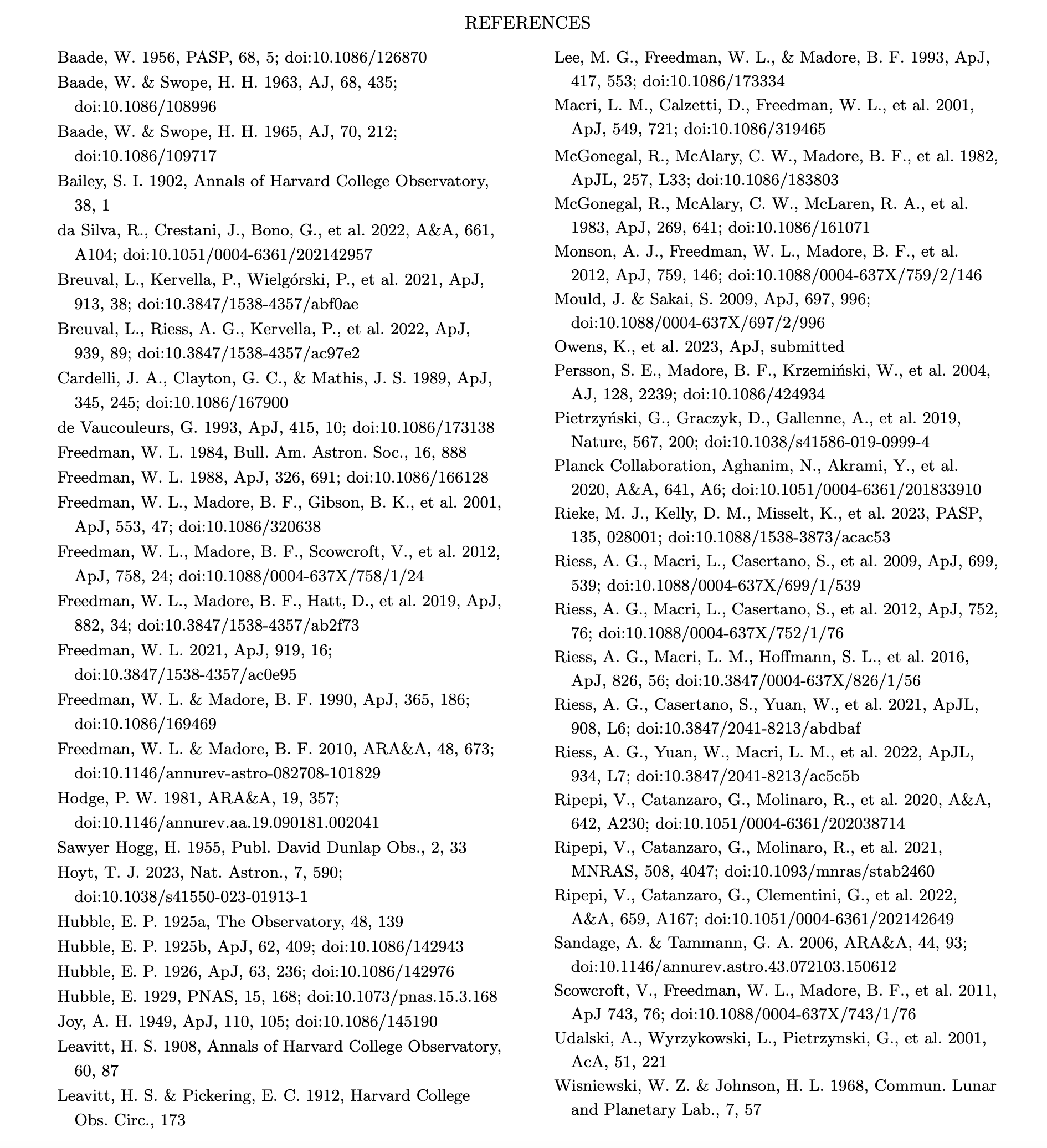}
\end{figure}

\end{document}